# Biomimetic fabrication and tunable wetting properties of three-dimensional hierarchical ZnO structures by combining soft lithography templated with lotus leaf and hydrothermal treatments

Shuxi Dai, Dianbo Zhang, Qing Shi, Xiao Han, Shujie Wang and Zuliang Du*

Three-dimensional hierarchical ZnO films with lotus-leaf-like micro/nano structures were successfully fabricated *via* a biomimetic route combining sol-gel technique, soft lithography and hydrothermal treatments. PDMS mold replicated from a fresh lotus leaf was used to imprint microscale pillar structures directly into a ZnO sol film. Hierarchical ZnO micro/nano structures were subsequently fabricated by a low-temperature hydrothermal growth of secondary ZnO nanorod arrays on the micro-structured ZnO film. The morphology and size of ZnO hierarchical micro/nano structures can be easily controlled by adjusting the hydrothermal reaction time. Wettability of hierarchical ZnO thin films was found to convert from superhydrophilicity to hydrophobicity after a low-surface-energy fluoroalkylsilane modification. Improved wetting properties from hydrophobic to superhydrophobic can be tuned by increasing the growth of ZnO nanorods structures.

## 1. Introduction

Many surfaces in nature such as the wings of insects and the leaves of plants are highly hydrophobic and exhibit self-cleaning property.[1-4] One of the most known examples with hydrophobic self-cleaning surfaces is lotus leaf.[2-4] Jiang et al. reported that the surface of lotus leaf is covered with dual-scale hierarchical structured protuberances.[2] Electron microscopy investigations show the primary structure of microscale papillae on the lotus leaf is covered with a secondary structure of nanoscale tubular wax crystalloids.[3] Further investigations showed that the epicuticular wax crystals contain predominantly $-CH_2$ groups as low surface energy materials.[4] Previous studies have shown that the combination of micrometer-scale and nanometer-scale roughness, along with a low surface energy material on the lotus surface leads to an apparent superhydrophobic property with the water contact angle > 150°, a low sliding angle and self-cleaning effect.[5] Inspired by the lotus leaf, a superhydrophobic surface can be prepared by a combination of enhancing the surface roughness with hierarchical micro/nano structures and lowering the surface energy through chemical surface modification.[6,7] To imitate the surface structure of a lotus leaf artificially, direct replication of lotus leaf is a facile and promising way.[7-10] Nanocasting based on soft lithography is a convenient and low-cost nanofabrication method for the patterning micro/nano structures. Sun et al. first fabricated a superhydrophobic poly(dimethylsiloxane) (PDMS) surface with micro/nano structures replicated from a natural lotus leaf as an original template.[8] Zhang et al. presented a simple technique to copy the lotus leaf structures onto poly(etherimide) film via a two-step phase-separation micromolding process at room temperature and normal pressure.[9] More hydrophobic polymer films such as polyvinylsiloxane and polycaprolactone were recently fabricated with the surface pattern replicated from the lotus leaf using pressure/electric-field-assisted micro/nanocasting method.[10,11]

Various artificial superhydrophobic surfaces of functional materials such as ZnO, $TiO_2$, carbon nanotubes, silica, and silver with unusual characteristics such as self-cleaning and anti-adhesion have been fabricated using different growth techniques such as colloidal lithography.[12,13] Among them, ZnO is one of the most attractive oxide semiconductor materials, due to its excellent electrical and optical properties. In particular, hierarchical ZnO nanostructures have attracted considerable attention owing to its applications to nanodevices such as light-emitting diodes,[14] UV light detector,[15] gas sensors,[16] solar cells,[17] etc. However, it still remains a big challenge to develop simple and reliable fabrication methods for ZnO hierarchical micro/nano architectures with controlled morphology, surface architectures and tunable wettability. ZnO tree-like hetero-nanostructures had been fabricated with the growth of ZnO nanorod arrays on primary silicon[18] and polycarbonate micropillars[19] obtained by top-down lithography methods. Shen et al. reported that hierarchical lotus-leaf-like ZnO micro-nanostructure film with strong adhesive force was fabricated by the growth of ZnO nanorods on the alumina pyramid substrate.[20] Recently, fresh lotus leaf was directly used as a natural substrate for the growth of ZnO nanowire arrays using a hydrothermal method.[21] A strong adhesive superhydrophobicity on this artificial/natural architecture was observed. It provides a new fabrication strategy for material engineering on real biological surfaces.

Herein we present a convenient technique for the fabrication of hierarchical lotus-leaf-like ZnO micro/nano structures. As shown in Figure 1, our method features a combination of direct imprinting of ZnO sol-gel film using a soft PDMS mold and subsequent hydrothermal growth of ZnO nanorod arrays. The PDMS mold was replicated from a lotus leaf. Then the microscale pillar structures were transferred to ZnO sol films with a modified soft lithography method under the applied pressure and heating. Secondary growth of well aligned arrays of ZnO nanorods were obtained on the primary micropillars of ZnO films using a simple low temperature hydrothermal treatment. The hierarchical lotus-leaf-like ZnO surface exhibited a tunable wettability from hydrophobic to superhydrophobic after the fluoroalkylsilane modification. To our knowledge, this is the first report of fabrication a homogenous ZnO hierarchical lotus-leaf-like film with superhydrophobic surface by replicating primary microstructures using a natural leaf as a template and subsequently growing secondary nanostructures. This fabrication process of ZnO lotus-leaf-like structures combined with simple sol-gel, casting and hydrothermal methods can be conveniently applied to the other functional metal oxide materials such as titanium dioxide and zirconium dioxide.

## 2. Experimental

### Material

All the regents and solvents were analytical grade from Aldrich and used without further purification. Deionized water (Millipore, USA) with a resistance of 18 MΩ cm$^{-1}$ was used. Silicon wafers were used as substrates and cleaned by a RCA standard cleaning method.

### Fabrication of ZnO micro-structures with imprint of PDMS mold

Fresh lotus leaves were collected from a pond in Henan University and cleaned with deionized water and dried with $N_2$ stream to remove dust particles. Fresh lotus leaf was cut into pieces with a size of 20 x 20 mm. A piece of lotus leaf was fixed onto a silicon wafer by a double-sided adhesive tape. Poly(dimethylsiloxane) (PDMS, Sylgard 184 kits, Dow Corning) was used to replicate the surface structures of lotus leaves and then act as a mold in the soft imprint lithography process. Sylgard 184 silicone elastomer base and curing agent were mixed (10:1 ratio by weight) for 2 min to obtain a homogeneous solution and degassed in a vacuum oven for 5 min to remove entrapped air bubbles. Then, PDMS elastomer mixture was poured on the fresh lotus leaf fixed in a polystyrene Petri dish. A relative lower temperature of 40 ℃ was used for the PDMS curing to prevent the heat damage of the natural lotus leaf surface structure. After 12 hours of solidification, the cured PDMS mold was peeled off from the lotus leaf and the edges were cut with a razor blade at the desired dimensions.

The imprint process and effect using PDMS mold on ZnO sol films was firstly tested on a commercial imprinter (NIL-2.5, Obducat, Sweden) according to our previous report.[22] Then, a simple and reproducible modified soft lithography method was used to imprint ZnO sol films. ZnO sol solution was prepared by dissolving 0.9 g $Zn(NO_3)_2·6H_2O$ and 0.9 g poly(acrylicacid) in 0.9 mL of 2-methoxyethanol and 0.9 mL of water under stirring. ZnO sol was spin-coated on a silicon wafer at 4000 rpm for 60 s and soft baked at 80 ℃ for 5 min. A PDMS mold replicated from a fresh lotus leaf was placed on the ZnO sol film and then imprinted into ZnO sol film under a specific pressure at 120 $^o$C for 300 s. After separation of the PDMS mold, the imprinted ZnO sol-gel film was calcined at 500 ℃ with 5 ℃/min heating rate in atmospheric ambient for 1 h to get crystallized ZnO film with microscale pillar structures originated from the lotus leaf.

### Growth and surface modification of ZnO Hierarchical micro/nano structures

The secondary growth of ZnO nanorod arrays on the crystallized ZnO microscale pillar structures were performed using a low temperature hydrothermal method. A 1:1 volume mixture 20 mM of $Zn(NO_3)_2$ $6H_2O$ and hexamethylenetetramine (HMTA, 99%, Aldrich) were transferred into a 50 mL Teflon-lined stainless steel autoclave and filled up to 80% volume. The microstructured ZnO film on silicon substrate was faced downward in the autoclave to avoid the precipitates on the ZnO surface. The hydrothermal synthesis was carried out at 80 ℃ for 2 - 24 h. Then the samples were taken out and rinsed thoroughly with deionized water and dried at 80 ℃. Finally, fluoroalkyl silane modified ZnO samples were obtained using vapor-phase deposition at 90 ℃ for 12 h in a three-neck flask with 2 ml toluene and 20 μl 1H,1H,2H,2H-perfluorodecyltriethoxysilane (PFDTES) inside. All silane modified samples were rinsed with toluene and dried under a nitrogen flow.

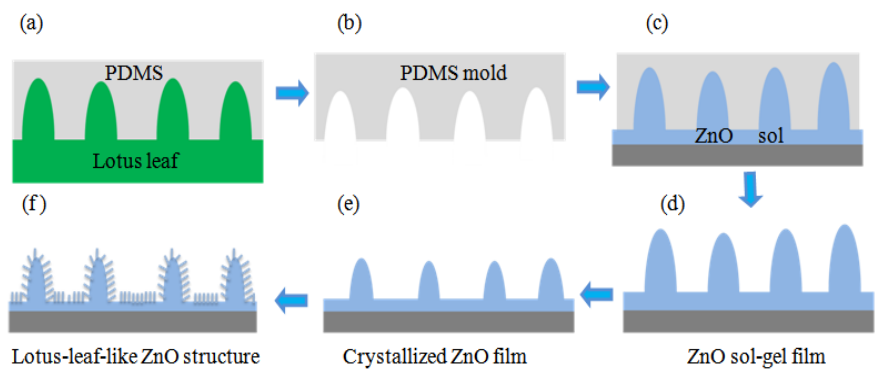

**Fig. 1** Schematic representation of the fabrication process of lotus-leaf-like ZnO structures: (a) PDMS replication of fresh lotus leaf, (b) replicated PDMS mold (c) soft imprint on ZnO sol film, (b) imprinted ZnO sol-gel film (e) crystallized ZnO film after calcination (f) lotus-leaf-like ZnO micro/nano structures after hydrothermal reaction.

**Morphology and structure characterizations**

The surface topographies of samples were investigated by a scanning electron microscopy (JSM-5600LV, JEOL LTD., Japan) at 20 kV. The crystal structure of films was identified by X-ray diffraction (XRD) using a Philips X-ray diffractometer (X'Pert Pro, Holland) with Cu Kα radiation. The static water contact angle was measured with an optical contact angle meter (Dropmaster 300, Kyowa Interface Science, Japan) at room temperature. The volume of the individual water droplet used for the static contact angle measurements was 4 μL. Contact angle values were obtained by averaging five measurements results on different surface areas of the sample.

## 2. Results and discussion

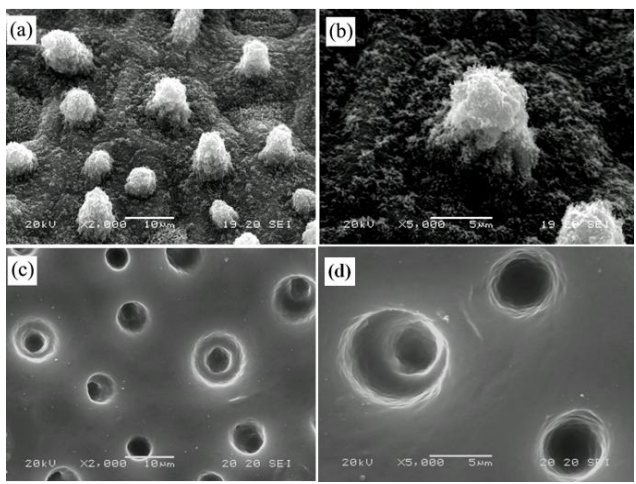

**Fig. 2** SEM images of (a,b) fresh lotus leaf surface composed of micro/nano scale binary structures and (c,d) negatively replicated PDMS mold.

Hierarchical micro/nano surface structure and composition of lotus leaves have been investigated by many researchers. Superhydrophobic behavior of lotus leaves is found to be a result of the rough hierarchical structure, as well as hydrophobic epicuticular wax layer present on the leaf surface. Fig. 2a-b presents the typical SEM images of a fresh lotus leaf surface composed of micro/nano scale binary structures. There are many microscale papillate hills distributed randomly on the surface of lotus leaf. Height of hills is in the range of 5 -10 μm, and diameter of the small hills and the distance between two hills are about 5 - 8 μm (Fig. 2b). In addition, the micro papillae and basal area of the lotus leaf randomly covered by branch-like nanoprotrusions of wax crystal with a diameter of about 100 - 150 nm. We measured the wettability of natural lotus leaf and obtained the static contact angle of about 160 ° corresponded to its superhydrophobic and low surface energy behavior, which is well in agreement with reports in the literature.[2-4]

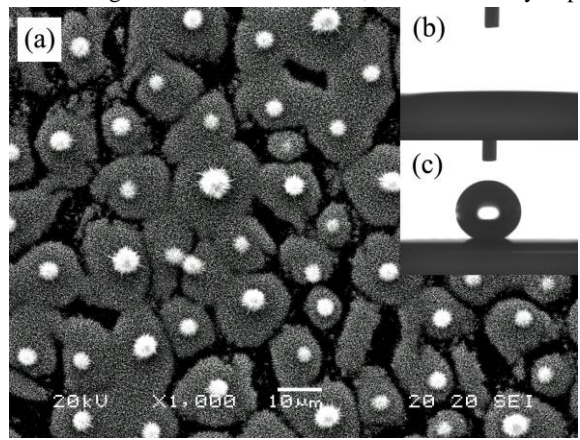

**Fig. 4** SEM image of (a) hierarchical ZnO micro/nanostructure prepared after 9h hydrothermal treatment. Insert shows the optical images of a water droplet on the (b) as-prepard sample and (c) sample after fluoroalkyl silane modification.

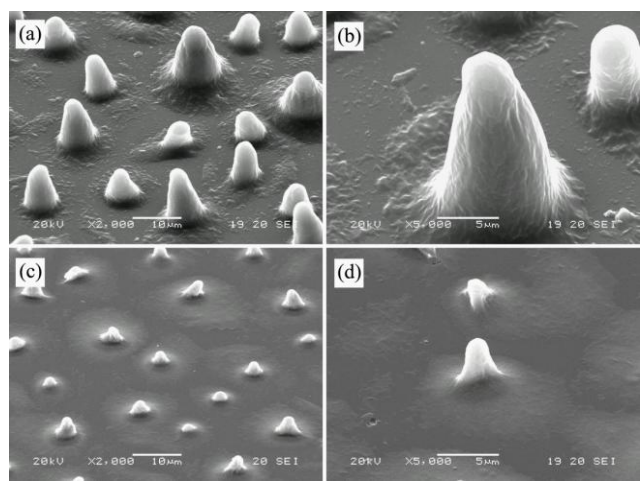

**Fig. 3** SEM images of (a,b) imprinted ZnO sol-gel film and (c,d) crystalline ZnO film after calcination at 500 °C for 2 h.

To replicate the unique surface structures of lotus leaf, liquid PDMS polymers were cast onto the fresh lotus leaf. Fig. 2c-d presents the SEM images of replicated PDMS after thermal curing at 40 °C and peeling off from the lotus leaf. The replicated surface shows lots of holes with diameter ranged from 3 to 10 μm on the large smooth area. The microcavities on the PDMS correspond to the negative structures of micro-sized pillars on the lotus leaf surface. The inner surface of the microcavities seems relatively smooth without obvious nanoscale structures. This indicates that the nanostructures of wax crystal on lotus leaf are hard to copy due to its small size, superhydrophobic and low surface energy behavior.

Positive lotus leaf replication on ZnO sol-gel film was obtained by imprinting of negative PDMS mold into the spin-coated ZnO sol film under specific applied pressure and heat using a modified soft lithography method. Fig. 3a-b presents the surface morphology of the imprinted ZnO sol-gel film. This positive replica shows a surface morphology of microscale papillary structure with a diameter of 3 – 11 μm and a height of 7 – 13 μm originated from lotus leaves. Sub-microscale textured structures were observed on the micrometer-scale pillars and basal area around micro pillars (Fig. 3b). It shows that the microstructures on the lotus leaf can be easily transferred onto ZnO sol-gel film with this easy casting and imprint method. The ZnO sol-gel layer with microscale papillary structure was then converted into crystalline ZnO film by calcination at 500 °C for 1 h. Fig. 3c-d shows the SEM images of crystallized ZnO film. The dimension of calcined micropillars structures was clearly reduced compared to that on the imprinted ZnO sol-gel film. Height and diameter of the pillars are in the range of 1 - 4 μm, and the distance between two pillars is 10 - 20 μm (Fig. 3d). This dimension decrease is due to the crystallization of zinc oxide and the thermal removal of organic residues at high temperature.

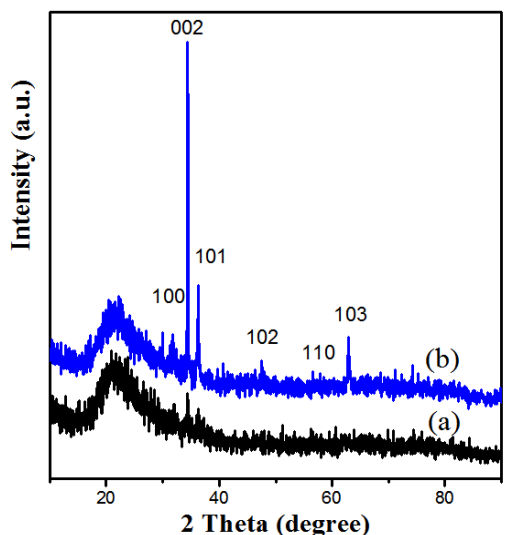

**Fig.5** XRD patterns of (a) crystalline ZnO film with micropillar structures after calcination at 500 °C for 2 h and (b) hierarchical ZnO micro/nano structures after hydrothermal treatments at 80 °C for 9 h.

To mimic the micro/nano structures of lotus leaf, a secondary growth of well aligned arrays of ZnO nanorods were obtained on the primary micropillars of ZnO films using a simple low temperature hydrothermal method. Three-dimensional ZnO micro/nano structure was fabricated after hydrothermal treatments of calcined ZnO film at 80 °C for several hours. Fig. 4a shows the typical SEM image of prickly ZnO micro/nanostructures prepared after 9h hydrothermal treatment. All the micropillars and basal areas around the pillars were covered by aligned nanorods arrays.

In addition, the nanorods grown on the top and side of micropillars are about 2-3 μm in length and 100 nm in diameter with a clear radial growth arrangement pointing toward the center of the micropillars. As we can see that the nanorods grown on the micropillars have relatively larger length and diameter than the nanorods grown on the flat basal areas. And the growth of nanorods shows a distinct boundary between two micropillar regions. Compared to the flat basal areas, the micropillars show more available surface contact to the reaction solution and the microstructured arrays may act as nucleation platforms for the preferential and rapid growth of nanorods on the top and side of micropillars. Further adjustment of the diameter and length of nanorods can be conducted by the change of hydrothermal growth parameters such as temperature, zinc salt and HMTA concentration.

Fig. 5 shows XRD patterns of crystallized ZnO films with micro-structures (a) and micro/nano structures (b) after hydrothermal treatments. Both XRD patterns correspond to typical diffraction peaks of polycrystalline ZnO with a hexagonal wurtzite structure (Zincite, JCPDS 36-1451). The low intensities and broadening of diffraction peaks in XRD pattern 5a indicated the formation of ZnO nanoparticles with small grain size. The polycrystalline ZnO film with small nanocrystallites can act as a good seed layer for the subsequent growth of ZnO nanostructures.[23] XRD pattern 5b of hierarchical ZnO micro/nanostructures exhibits a strong (002) diffraction peak, indicating that the ZnO nanorods are highly c-axis oriented with high purity of the hexagonal ZnO phase and good crystallinity.

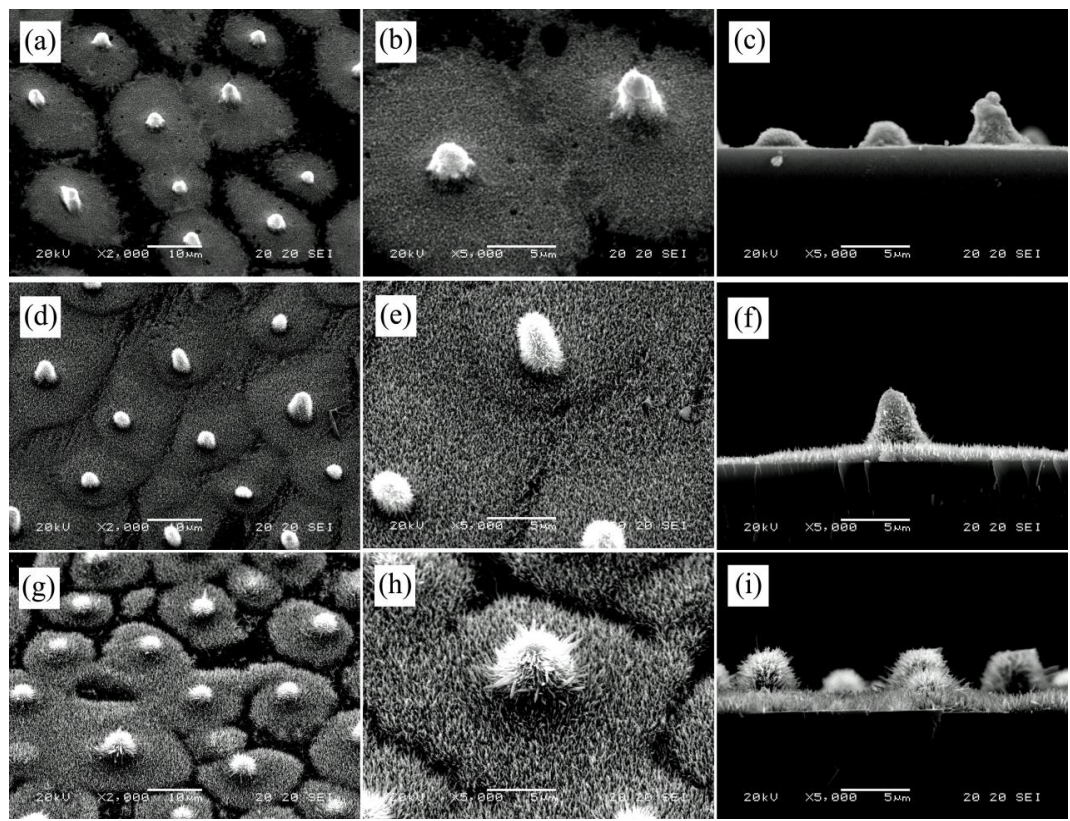

**Fig. 6** Typical SEM top-view and cross-section images of the as-prepared lotus-leaf-like hierarchical ZnO surfaces obtained at a series of hydrothermal growth times for (a, b, c) 3h, (d, e, f) 6h and (g, h, i) 9h.

Water CA measurements were performed on the prepared hierarchical ZnO micro/nanostructures. Fig. 4b shows the optical image of water droplet behavior on the fresh ZnO micro/nanostructures prepared after hydrothermal treatment for 9 h. The surface was found to be superhydrophilic with a contact angle about 3° (Fig. 4b). Bando et al. reported that the ZnO nanorods arrays prepared by the hydrothermal method presented surface hydroxyl groups which were thermally stable up to the temperature of 400 °C.[24] Recently, Yong et al. demonstrated similar superhydrophilic surface of the as-grown ZnO nanorod arrays with a good antifogging property.[25] The superhydrophilicity of fresh hierarchical ZnO surface indicates that only the construction of hierarchical micro/nano structures is not enough to obtain the lotus-leaf-like hydrophobic wettability. Low surface energy materials such as silanes and fatty acids are essential to absorb onto the ZnO hierarchical surface and worked as the function of wax nanocrystals on the lotus leaf. In our work, a self-assembled monolayer of perfluorodecyltriethoxysilane was deposited on the hydrothermally prepared ZnO micro/nanostructures. After treatment with fluoroalkylsilane, the hierarchical ZnO micro/nanostructures exhibited a dramatic conversion from superhydrophilicity to superhydrophobicity with a water CA changed from 3° to 152° (Fig. 4c). For comparison, a fluoroalkylsilane modified flat ZnO surface showed only hydrophobicity with a CA of about 110°. The above CA measurements showed the fluoroalkylsilane treatments play more roles in the hydrophilicity to hydrophobicity transformation by preventing direct contact with ZnO surface. And the hierarchical micro/nanostructures contribute more to the transformation from hydrophobicity to superhydrophobicity after surface modification with fluoroalkylsilane.

To investigate the influence of the nanoscale structures on the wettability of hierarchical ZnO micro/nano surface, samples obtained at different growth stages of the hydrothermal reaction were examined by SEM observations and CA measurements. Figure 6 presents the morphological evolution of the as-prepared hierarchical ZnO films obtained at a series of hydrothermal growth times. It was shown that the hierarchical ZnO films exhibited surface roughness at two length scales prepared by the consecutive formation of ZnO micropillars and nanorod arrays. ZnO pillars with microscale roughness were fabricated with soft lithography and subsequent calcination. Then, nanoscale roughness was achieved by the hydrothermal growth of ZnO nanorods on the crystallized micro-structured ZnO surface. Increase in length and diameter of ZnO nanorods grown on the crystallized microstructured ZnO surface were observed with the increased reaction time. As shown in the cross-section images (Fig. 6c, 6f and 6i), the surface roughness of the hierarchical ZnO micro/nano surface increased evidently with the growth of ZnO nanorod arrays.

Fig. 6a-c show the morphology of ZnO film obtained at an early reaction stage of 3 h at 80 °C. The ZnO surface presents a similar microscale pillars as the calcined ZnO film due to the short reaction time. High magnitude SEM observations show tiny ZnO particles and a few nanorods with diameter less that 50 nm were obtained on the micropillars and the flat basal area around the pillars. When the reaction time was increased to 6 h, lotus-leaf-like hierarchical structures consisting of micropillars and nanorods were obtained. A side view shown in Fig. 6f suggests nearly vertically oriented arrays of ZnO nanorods (about 500 nm in length) densely covered the surface of micropillars and flat surrounding areas. After 9 h of reaction, urchin-like ZnO micro/nano structures were clearly observed in the Fig. 6 g-i. A large number of longer nanorods grown from the center of micropillars and covered the surrounding areas on the substrate with diameters and lengths of about 100 nm and 2 μm. It indicted the larger surface area of ZnO microstructures exposed to zinc precursor solution increased the nuclei supersaturation and the ZnO nuclei on the micropillars could grow faster to form longer nanorods of ZnO than those on the flat areas. ZnO nanorods around the micropillars with faster fast growth orientations lead to urchin-like hierarchical structures. We also performed the hydrothermal synthesis for longer reaction time such as 12 h and 24 h. The results showed that the morphologies of the micro/nano structures had not changed much and only the dimensions of the nanorods grown on the flat area were slightly increased. This structural change of ZnO micro/nano structures indicated that the micropillars had lost their advantages in the fast growth of ZnO nanorods at the final stage due to the low concentration left or run out of the zinc precursor solution.

CA measurements were performed on the ZnO micro/nano structures after surface modification with fluoroalkylsilane. Fig. 7 shows digital optical images of water droplets on the fluoroalkylsilane modified ZnO micro/nano structures with hydrothermal growth for 3 h, 6 h, 9 h and 12 h, respectively. The corresponding static CA values in Fig 7 a-d are 112°, 140°, 153° and 158°, respectively. Combined CA data in Fig. 7 and the SEM observations shown in Fig. 6, it showed the morphology and dimension of the nanostructures had essential influences on the hydrophobic properties of hierarchical ZnO surface. Fig. 7a shows the ZnO micropillars structures had a CA value of 112 ° after hydrothermal growth for 3 h without the formation of obviously secondary nanorods structures. After 6 h growth, the micropillars and surrounding surface were fully covered with nanorods arrays with a length of about 500 nm, which gave rise to a dramatic increase of CA value to 140 °(Fig. 7b). The urchin-like ZnO structures with nanorods of 2 μm in length growth exhibited a superhydrophobic property with contact angle of 153 ° (Fig. 7c) after 9 h hydrothermal treatments. When the reaction time was increased to 12 h, the nanostructures on the ZnO microstructured surface had no obvious change. A slight increase of CA value to 158 ° was obtained due to the further growth of ZnO nanostructures on the flat regions of the ZnO surface. CA measurements showed that the superhydrophobic surface formed only after 9 h hydrothermal growth with the length scale of 2 μm and width scale of 100 nm for nanorod structures. It indicates that the replication of primary micropillar structures and the subsequent surface modification of low energy fluoroalkylsilane are responsible for the wettability conversion from hydrophobicity to hydrophobic property. The nanorods structures hydrothermally grown the micro-scale structures make it possible to get a superhydrophobic surface. And it shows the control of the diameter, length, and density of the secondary ZnO nanorods is essential to reach superhydrophobicity. When a water droplet was placed on the modified hierarchical ZnO micro/nano structures, the fluoroalkylsilane modified micropillars/nanorods surfaces totally prevent the penetration of water into the interspace of nanorods to give rise to a hydrophobic property. The secondary nanorods structures with certain length and width (length scale of 2 μm and width scale of 100 nm) enables air to be trapped more easily under water droplets and thus most parts of the water droplet sit on the air layer, resulting in a superhydrophobic surface with a CA value larger than 150º.

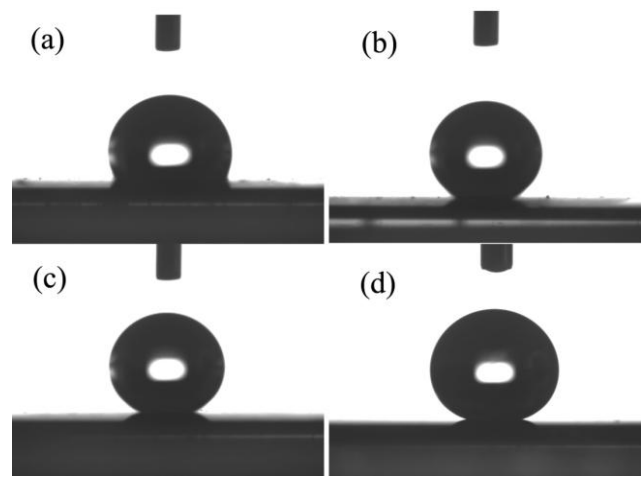

**Fig. 7** Optical images of water contact angles of fluoroalkylsilane modified ZnO micro/nano structures with hydrothermal growth for (a) 3h, (b) 6h, (c) 9h and (d) 12h.

To gain more understanding of the effect of secondary ZnO nanostructures on the hydrophobic property, dynamic CA measurements were conducted on the fluoroalkylsilane modified ZnO micro/nano structures obtained from different growth stages. Fig. 8 shows the advancing CAs and receding CAs of ZnO micro/nano structures as a function of the hydrothermal reaction time. The difference between advancing and receding CA is the contact angle hysteresis (CAH). As shown in Fig. 8, the dynamic water CA, both advancing and receding CAs, of the fluoroalkylsilane modified ZnO micro/nano surfaces increased monotonically with increasing hydrothermal growth time. For the early stage of hydrothermal growth for 3 h, the ZnO surface had the advancing CA value of 115 ° with only micropillars structures without secondary nanorods structures (see Fig. 6a-c). The advancing CA increased dramatic to 147 ° with the formation of secondary nanorods covered on the micropillars and surrounding flat surface (Fig. 6d-f). Then the ZnO micro/nano surfaces obtained with growth time of 9 - 24 h had advancing CAs greater than 156 ° and kept stable because the length and width of nanorods on the micropillars kept almost no change after the 9 h growth. Finally the CA value reached the maximum of 160 °after 24 h growth. The magnitude of CAH of ZnO samples obtained from 3 h to 9 h increased from 16 to 23 with increasing length and width of secondary ZnO nanorods. After 9 h growth, the ZnO surface exhibited superhydrophobic property with a certain length of ZnO secondary nanorods structures. And the difference of advancing CA and receding CA became smaller with the decrease of CAH value after 9 h growth due to the gradual stop of ZnO nanorods growth.

## Conclusions

In summary, we have demonstrated a biomimetic route for the fabrication three-dimensional hierarchical ZnO film with lotus-leaf-like micro/nano structures with tunable wetting property. The fabrication process features a combination of direct imprinting of ZnO sol film using a lotus leaf template and subsequent hydrothermal growth of ZnO nanorod arrays. The Secondary growth of ZnO nanorod arrays on the primary micropillars of ZnO films could significantly improve the superhydrophobic property. This fabrication process of ZnO lotus-leaf-like structures combined with simple sol-gel, soft lithography and hydrothermal methods. Hierarchical ZnO and other metal oxide lotus-leaf-like structures can be fabricated via this large-area PDMS casting and low cost solution processed hydrothermal method.

## Acknowledgements

This work was supported by the Program for Changjiang Scholars and Innovative Research Team in University (No. PCSIRT1126), National Natural Science Foundation of China (No. 20903034 and 11274093), Young Backbone Teacher Cultivating Project of Henan University and SRF for ROCS, SEM.

## Notes and references

*Key Laboratory for Special Functional Materials of Ministry of Education, Henan University, Kaifeng 475004, P.R. China.*
*E-mail: zld@henu.edu.cn*

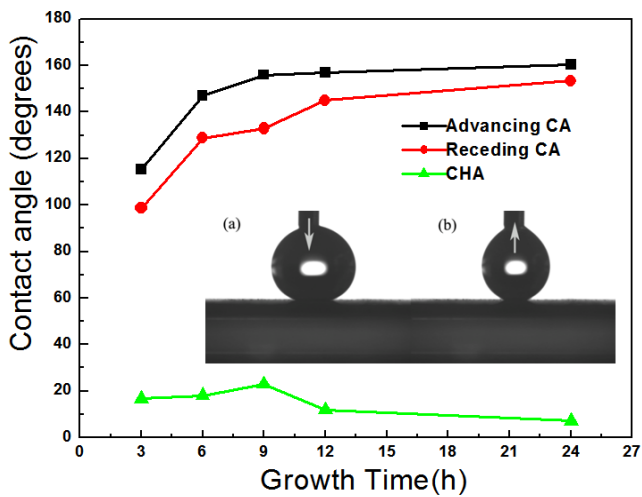

**Fig. 8** Advancing CAs, receding CAs and contact angle hysteresis (CHA) as a function of the hydrothermal reaction time for the surface modified ZnO micro/nano structures. Insert shows typical optical images of (a) advancing CA and (b) receding CA.